\documentclass[a4paper]{article}

\usepackage{INTERSPEECH2021}
\usepackage{adjustbox}
\usepackage{multirow}
\usepackage{multicol}
\usepackage{amssymb}
\usepackage{subcaption}
\usepackage{url}
\newlength{\bibitemsep}
\newlength{\bibparskip}
\let\oldthebibliography\thebibliography
\renewcommand\thebibliography[1]{%
  \oldthebibliography{#1}%
  \setlength{\parskip}{\bibitemsep}%
  \setlength{\itemsep}{\bibparskip}%
  \footnotesize
}

\usepackage{array}
\newcolumntype{L}[1]{>{\raggedright\let\newline\\\arraybackslash\hspace{0pt}}m{#1}}

\title{Speech based Depression Severity Level Classification Using a Multi-Stage Dilated CNN-LSTM Model}
\name{Nadee Seneviratne$^1$, Carol Espy-Wilson$^1$}
\address{
  $^1$University of Maryland - College Park, USA}
\email{nadee@umd.edu, espy@umd.edu}

\begin{document}

\maketitle
\begin{abstract}
Speech based depression classification has gained immense popularity over the recent years. However, most of the classification studies have focused on binary classification to distinguish depressed subjects from non-depressed subjects. In this paper, we formulate the depression classification task as a severity level classification problem to provide more granularity to the classification outcomes. We use articulatory coordination features (ACFs) developed to capture the changes of neuromotor coordination that happens as a result of psychomotor slowing, a necessary feature of Major Depressive Disorder. The ACFs derived from the vocal tract variables (TVs) are used to train a dilated Convolutional Neural Network based depression classification model to obtain segment-level predictions. Then, we propose a Recurrent Neural Network based approach to obtain session-level predictions from segment-level predictions. We show that strengths of the segment-wise classifier are amplified when a session-wise classifier is trained on embeddings obtained from it.
 The model trained on ACFs derived from TVs show relative improvement of 27.47\% in Unweighted Average Recall (UAR) at the session-level classification task, compared to the ACFs derived from Mel Frequency Cepstral Coefficients (MFCCs).

\end{abstract}

\section{Introduction}
With more than 264 million people suffering worldwide \cite{WHO2020}, Major Depressive Disorder (MDD) is one the most critical mental health disorders that affects the quality of life. MDD can even lead to suicidality and that urges the requirement of timely diagnosis and prompt treatments. 
Previous studies have shown that vocal biomarkers developed using prosodic, source, and spectral features \cite{CUMMINS201510, Scherer2013,Cummins2013b} can be very useful in depression detection and severity prediction.

Articulatory Coordination Features (ACFs) have yielded successful results in distinguishing depressed speech from non-depressed speech by quantifying the changes in timing of speech gestures \cite{Williamson2014,WILLIAMSON2019, Espy-Wilson2019, Seneviratne2020}. These changes in articulatory coordination happens as a result of neurological condition called psychomotor slowing, a necessary feature of MDD that is used to evaluate the severity of MDD \cite{Whitwell1937, ManualMentalDisOrd, WIDLOCHER198327}. Previously, the correlation structure of the formants or MFCCs were used as a proxy for articulatory coordination to derive indirect ACFs which showed promise in the depression detection task \cite{WILLIAMSON2019}. Authors of this paper showed in their previous work, that by using Vocal Tract Variables (TVs) as a direct measure of articulation to quantify changes in the way speech is produced by depressed and non-depressed subjects can yield significantly better results in depression detection task \cite{Espy-Wilson2019, Seneviratne2020}. 
In recent work, the authors applied the time-delay embedded correlation matrix derived from TVs as ACFs to train a generalized deep learning based model for the first time with speech data sourced from two depression databases with different characteristics . It was shown that TV based ACFs show promise as a robust set of features for depression by generalizing well across the two databases \cite{seneviratne2020deep}. 

Most previous studies on depression classification focused on detecting whether a subject is depressed or not \cite{Low2011, Cummins2011, Vlasenko2017, Ma2016} or detecting high or low depression \cite{Scherer2013, Helfer2013}. A very few studies have looked into performing classifications across more than just 2 classes \cite{Trevino2011}, but not using deep learning based models. Hence we extend our work to perform a depression severity level classification across 3 classes (normal, moderate, and severe) using TV based ACFs. This helps to identify those who are at critical stages with severe depression, allowing to prioritize the allocation of limited resources. Then we propose a multi-stage model to perform session-wise classifications using segment-level classifications. We show that this technique can result in significant improvements in final classifications than training models using features extracted directly from full audio recordings by helping to avoid overfitting issues due to high dimensionality of the input features and low amount of training samples. We perform experiments using multiple feature sets (MFCCs, Formants, openSMILE features) to compare against the results of TV based ACFs. 

The paper is organized as follows: section 2 explains the methodology involving feature extraction and model architectures. Section 3 presents the experiments conducted and results obtained. Section 4 analyses the results in detail with potential future directions.

\section{Method}
\subsection{Depression Databases}
We use speech data from two depression databases \cite{MUNDT2007, MUNDT2012} (Table \ref{tab:dep_databases}).
Two clinician (CL)-rated depression assessment scales: Hamilton Depression Rating Scale (HAMD) and Quick Inventory of Depressive Symptomatology (QIDS) were provided which were used to define the severity levels of depression (Table \ref{tab:dep_severity}).  For the 3-class severity level classification task, data in levels 4-5 and 2-3 for was combined for classes `severe' and `moderate' respectively. Data in level 1 was used for `normal' class.

\vspace{-10pt}
\begin{table}[h]
    \centering
    \scriptsize
    \caption{\textit{Severity level definitions of MDD assessment scales}}
    \label{tab:dep_severity}
    \begin{adjustbox}{max width = \columnwidth}
        \begin{tabular}{ccc}
            \toprule
            \textbf{Severity Level} & \textbf{HAMD} & \textbf{QIDS} \\ \midrule
            1. Normal & 0 – 7 & 0 - 5 \\ 
            2. Mild & 8 - 13 & 6 - 10 \\ 
            3. Moderate & 14 - 18 & 11 - 15 \\
            4. Severe & 19 - 22 & 16 - 20 \\ 
            5. Very Severe & 23 - 52 & 21 - 27 \\ \bottomrule
        \end{tabular}
    \end{adjustbox}
    \vspace{-10pt}
\end{table}


\begin{table}[h]
    \centering
    \caption{\textit{Details of Depression Databases}}
    \label{tab:dep_databases}
    \begin{adjustbox}{max width = \columnwidth}
        \begin{tabular}{ccc}
        \toprule
        \multicolumn{1}{c}{\textbf{Database}} & \multicolumn{1}{c}{\textbf{MD-1 \cite{MUNDT2007}}} & \multicolumn{1}{c}{\textbf{MD-2 \cite{MUNDT2012}}} \\ \midrule
        Longitudinal & 6 Weeks & 4 Weeks \\ 
        Study Type & Observational & Clinical trial \\
        \# Subjects & 20 F, 15 M & 104 F, 61 M \\ 
        Demography & 31 Caucasian & 125 Caucasian \\
         & 1 African American & 26 African American \\
         & 1 Bi-racial & 4 Asian \\
         & 1 Greek, 1 Hispanic & 10 Other \\ 
        Assessment & HAMD-CL: Bi-weekly & HAMD-CL, QIDS-CL: Weeks 1,2,4 \\ 
        FS Lengths & \multicolumn{1}{c}{Min: 2.5s, Max: 156.8s} & \multicolumn{1}{c}{Min: 2.6s, Max: 181.2s} \\ 
        Recording Type & \multicolumn{2}{c}{Interactive Voice Response Technology (8kHz)} \\ \bottomrule
        \end{tabular}
    \end{adjustbox}
    \vspace{-5pt}
\end{table}

\vspace*{-2pt}
In this study, we used recordings of free speech (FS) where patients describe how they feel emotionally, physically and their ability to function in each session. 
In MD-2, depression assessment scores were provided to only 105 subjects. Due to the availability of two CL-rated scores in MD-2, only the speech samples where both the scores belong to the same severity level were used.

\subsection{Estimation of Vocal Tract Variables (TVs)}
\emph{\textbf{Acoustic-to-Articulatory Speech Inversion(SI): }}
We use the TVs estimated by a speaker independent, deep neural network (DNN) based SI system to estimate the low-level feature vectors used to compute ACFs. Articulatory Phonology (AP) \cite{Browman1992} views speech as a constellation of overlapping gestures. These gestures are discrete action units whose activation results in constriction formation or release by five distinct constrictors along the vocal tract: lips, tongue tip, tongue body, velum, and glottis). The vocal tract variables (TVs) are defined by the constriction degree and location of these five constrictors. The SI system computes the trajectory of the TVs that represent constriction location and degree of articulators located along the vocal tract \cite{Sivaraman2019}. The six TVs estimated by the SI system are – Lip Aperture, Lip Protrusion, Tongue Body Constriction Location, Tongue Body Constriction Degree, Tongue Tip Constriction Location and Tongue Tip Constriction Degree. For detailed information about the SI system, the reader is referred to \cite{Sivaraman2019}.

\emph{\textbf{Glottal TV Estimation: }}
To achieve a complete representation of TVs described by the AP, TVs corresponding to the glottal state should also be included. The same DNN based SI system could not be trained due to unavailability of ground-truth articulatory data as a result of difficulty in placing sensors near the glottis. Therefore, we augment the previous 6 TVs with the  periodicity and aperiodicity measures obtained from the Aperiodicity, Periodicity and Pitch detector developed in \cite{Deshmukh2005} which are used as glottal TVs. In \cite{Seneviratne2020}, these glottal parameters boosted the classification accuracies for depressed and non-depressed speech classification by about 8\%.

\subsection{Other Feature Extraction}
In order to compare the performance of the model trained using TV based `direct' ACFs, we also trained models with the same architecture using widely used MFCCs and Formants based ACFs which are considered as proxies for actual articulatory coordination. 12 MFCC time series were extracted by using an analysis window of 20 ms with a 10 ms frame shift (1\textsuperscript{st} MFCC coefficient was discarded). The first three formant frequencies were extracted using Praat \cite{Boersma2001}. Settings were: tracking 5 formants, 5500 Hz maximum formant, window length of 25ms and time step of 10ms.

To train a baseline model for comparison purposes, the extended Geneva Minimalistic Acoustic Parameter Set (eGeMAPS) was extracted using the openSMILE toolkit \cite{opensmile2010}. This 23-dimensional feature set consists of prosodic features like pitch, loudness, jitter, shimmer and other spectral parameters. The motivation behind using this feature set as the baseline case is because previous work on speech based depression recognition has shown that prosodic features like reduced speaking intensity, reduced pitch range and slower speech are characteristics of depressed speech \cite{Hollien1980}. These features were computed with a window size of 20ms with an overlap of 10ms.

\subsection{Articulatory Coordination Features}
We use the channel-delay correlation matrix proposed in \cite{Huang2020} as the ACFs for this work. This matrix overcomes the limitations found in the conventional approach \cite{WILLIAMSON2019} such as repetitive sampling and matrix discontinuities at the borders of adjacent sub-matrices.

For an $M$-channel feature vector $\mathbf{X}$, the delayed correlations ($r_{i,j}^d$) between $i^{th}$ channel $\mathbf{x_i}$ and $j^{th}$ channel $\mathbf{x_{j}}$ delayed by $d$ frames, are computed as:
\vspace{-5pt}
\begin{equation}
    r_{i,j}^d = \frac{\sum_{t=0}^{N-d-1}x_i[t]x_j[t+d]}{N-|d|}
\end{equation}
where N is the length of the channels.
The correlation vector for each pair of channels with delays $d \in [0,D]$ frames will be constructed as follows:
\vspace{-5pt}
\begin{equation}
    R_{i,j} = \begin{bmatrix}r_{i,j}^0, & r_{i,j}^1, & \dots & r_{i,j}^D \end{bmatrix}^T \in \mathbb{R}^{1\times (D+1)}
\end{equation}
The delayed auto-correlations and cross-correlations are stacked to construct the channel-delay correlation matrix:
\begin{equation}
    \mathit{\widetilde{R}_{ACF}} =  \begin{bmatrix}R_{1,1} & \dots & R_{i,j} & \dots & R_{M,M}\end{bmatrix}^T \in \mathbb{R}^{M^2\times (D+1)}
\end{equation}

It is important to note that the $\mathit{\widetilde{R}_{ACF}}$ matrix contains every correlation only once. With this representation, information pertaining to multiple delay scales can be incorporated into the model by using dilated CNN layers with corresponding dilation factors while maintaining a low input dimensionality. Each $R_{i,j}$ will be processed as a separate input channel in the CNN model, and thereby overcoming discontinuities. Before computing the correlation matrices, feature vectors (TVs) were standardized individually.

\begin{figure*}[t]
    \centering
    \includegraphics[width=\textwidth]{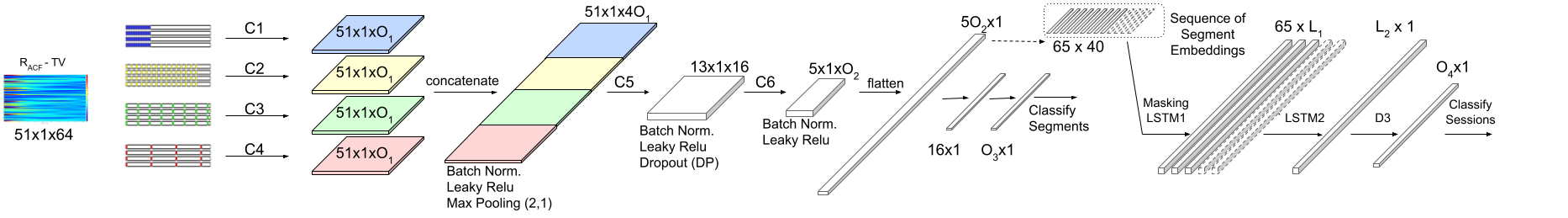}
    \caption{\textit{Dilated CNN-LSTM Architecture for Session-Level Classification}}
    \label{fig:dnn_model}
    \vspace{-15pt}
\end{figure*}

\subsection{Baseline Model}
We trained a CNN using openSMILE features to be used as a baseline model for the session-wise classification. The input is passed through two sequential convolutional layers ($Conv1$, $Conv2$). Each convolutional layer is followed by batch normalization, leaky ReLU activation, dropouts ($D_1$, $D_2$) and a max-pooling layer ($MP1$, $MP2$). The output from the second max-pooling layer is flattened and passed through a fully connected layer ($Dense1$) with ReLU activation with $l_2$ regularization of $\lambda=0.01$  to perform 3-class classification in the output layer.- 

The output of the $Dense1$ dense layer is extracted and used as the input to the LSTM model described in section \ref{sec:cnn-lstm-arch} to obtain session-wise classifications.

\subsection{Multi-Stage Dilated CNN-LSTM Model}
\label{sec:cnn-lstm-arch}
A multi-stage approach to obtain session-wise classification predictions from segment-wise predictions is motivated by the fact that the original number of speech recordings were insufficient to train a deep neural network with high-dimensional input features (since each recording serves as a training sample). To overcome this drawback, we propose a two-stage neural network architecture. 

In the first stage, a \textbf{dilated CNN} proposed in \cite{Huang2020} was trained using the ACFs to predict the segment-wise classifications (Fig. \ref{fig:dnn_model}). The input $\mathit{\widetilde{R}_{ACF}}$ is fed into four parallel convolutional layers ($C1,C2,C3,C4$) with different dilation rates $n=\{1,3,7,15\}$ and a kernel size of $(15,1)$ which resembles the multiple delay scales in the conventional approach. The outputs of these four parallel layers are concatenated and then passed through two sequential convolutional layers $(C5,C6)$. This output is flattened and passed through two fully connected (dense) layers $(D1,D2)$ to perform segment-level classification in the output layer. All convolutional layers used LeakyReLU activation, whereas dense layers used ReLU activation with $l_2$ regularization of $\lambda=0.01$. Batch Normalization, dropouts, and max-pooling layers were added as shown in the Fig. \ref{fig:dnn_model}. The weight sharing nature of CNNs handles the high dimensional correlation matrices with a low number trainable parameters.

The output of the first Dense layer ($D1$) 
is passed as input to the second stage. Since the segments of the sessions can be represented in sequential order, these embeddings are passed through an \textbf{LSTM based RNN} model to perform the session-level classification. The input is first passed through two LSTM layers ($LSTM1$ with $L_1$ units, returning sequence outputs and $LSTM2$ with $L_2$ units), followed by a Dense layer ($D3$) with ReLU activation and $O_4$ output units. 
Finally, the output layer with Softmax activation performs the session-level classification. Recurrent dropout probabilities of $DP_1$ and $DP_2$ are applied to the two LSTM layers respectively.

\section{Experiments \& Results}
\subsection{Dataset Preparation}

Originally there were 472 (35 speakers) and 753 (105 speakers) FS recordings from MD-1 and MD-2 respectively. The 140 speakers were divided into train / validation / test splits ($60:20:20$) preserving a similar class distribution in each split and ensuring that there are no speaker overlaps. For the models trained on TV, MFCC and Formant based ACFs, to increase the number of samples in order to train a deep neural model and to make the model resilient to translation, we segmented the audio recordings in the train and validation splits that are longer than 20s into segments of 20s with a shift of 5s. Recordings with duration less than 10s were discarded and other shorter recordings (between 10s-20s) were used as they were. Table \ref{tab:durations} summarizes the amount of speech data available after the segmentation. Test data was split into non-overlapping equal-length segments close to 20s. For the baseline model trained on openSMILE features, all these audio segments were truncated at 10s (minimum length of the available audio segments) to have fixed sized inputs to the CNN. Before extracting the low-level features, all audio segments were normalized to have a maximum absolute value of 1.
\begin{table}[h]
    \centering
    \scriptsize
    \caption{\textit{Duration of Available Data for the Dilated CNN model in hours (\# segments)}}
    \label{tab:durations}
    \begin{adjustbox}{max width = \columnwidth}
        \begin{tabular}{llll}
        \toprule
        \textbf{Database} & \textbf{Severe} & \textbf{Moderate} & \textbf{Normal} \\ \midrule
        MD-1              & 4.13 (763)      & 6.57 (1215)       & 2.17 (396)      \\
        MD-2              & 8.8 (1635)      & 5.63 (1044)       & 0.87 (164)      \\ \bottomrule
        \end{tabular}
    \end{adjustbox}
    \vspace{5pt}
\end{table}

\subsection{Model Training}

\emph{\textbf{Baseline Model: }}
Kernel size and number of output filters for $Conv1$ is 8 and 256 respectively and for $Conv2$ 8 and 128 respectively. The pool size of $MP1$ and $MP2$ is 8. $D_1$ and $D_2$ were tuned to 0.5 and 0.7 respectively. $Dense1$ has 64 units. These hyper-parameters were tuned using a grid search. A learning of $2e-5$ was used. Each dimension of each input openSMILE feature vector was individually standardized.

\emph{\textbf{Multi-Stage Dilated CNN-LSTM Model: }}
Using a maximum delay $D$ as 50 (empirically determined), the dilated CNN model was trained using a learning rate of $2e-5$. For $C5$, kernel size was $(3,1)$ with a stride of $2$ and 16 output filters were used. For $C1-C5$, `same' padding was used and for $C6$ `valid' padding was used. All input ACFs were standardized using the mean and the standard deviation of the training data. The LSTM model was trained using an adaptive learning rate starting from $2e-4$ and it was decayed by 50\% every 10 epochs until it reached $2e-5$. LSTM model hyper-parameters can be found in Table \ref{tab:lstm-params}.

The models were optimized using an Adam Optimizer for the Categorical Cross Entropy loss. The models were trained with an early stopping criteria based on validation loss (patience 15 epochs) for a maximum of 300 epochs. Batch size of 128 was used. To address the class imbalance issue, class weights were assigned to both training and validation splits during the training process to both the models. To evaluate the performance of the model, overall accuracy, Unweighted Average Recall (UAR), and F1 scores were used. Grid search was performed to tune the hyper-parameters of the dilated CNN model using the ranges in Table \ref{tab:hyperparam}. 

\begin{table}[h]
    \centering
    \small
    \caption{\textit{Grid Search Parameters for Best Models}}
    \label{tab:hyperparam}
    \begin{adjustbox}{max width = \columnwidth}
    \begin{tabular}{lccccc}
    \toprule
     & \textbf{C1-C4 Filter} & \textbf{C6 Filter} & \textbf{C6 Kernel} & \textbf{D2 Output} & \textbf{Dropout} \\
     & \textbf{Outputs ($O_1$)} & \textbf{Output ($O_2$)} & \textbf{($K_1$)} & \textbf{Size ($O_3$)} & \textbf{Prob. ($DP$)} \\ \midrule
    \textbf{Range} & \{16,32\} & \{8,16\} & \{(3,1), (4,1)\} & \{8,16\} & \{0.4,0.5\} \\ 
    \textbf{TV} & 16 & 8 & (4,1) & 16 & 0.5 \\
    \textbf{MFCC} & 32 & 16 & (3,1) & 8 & 0.5 \\ 
    \textbf{Formants} & 32 & 8 & (4,1) & 16 & 0.4 \\ \bottomrule
    \end{tabular}
    \end{adjustbox}
\end{table}
\vspace{-15pt}

\begin{table}[h]
\centering
\small
\caption{LSTM Based RNN Model Parameters}
\label{tab:lstm-params}
 \begin{adjustbox}{max width = \columnwidth}
\begin{tabular}{llllll}
\toprule
         & \textbf{LSTM1} & \textbf{LSTM2} & \textbf{LSTM1} & \textbf{LSTM2 } & \textbf{D3 Output}  \\
          & \textbf{Units ($L_1$)} & \textbf{Units ($L_2$)} & \textbf{Dropouts ($DP_1$)} & \textbf{Dropouts ($DP_2$)} & \textbf{Size $O_4$}  \\ \midrule
\textbf{TV}       & 64       & 64       & 0.4        & 0.3       & 32        \\
\textbf{MFCC}     & 128      & 64       & 0.6        & 0.4       & 64        \\
\textbf{Formants} & 128      & 64       & 0.7        & 0.7       & 16     \\ \bottomrule  
\end{tabular}
\end{adjustbox}
\end{table}

\vspace*{-15pt}

\subsection{Segment-Level Classification Results}
\label{sec:seg-results}

Table \ref{tab:overall-segment} includes the accuracy and UAR results for the models trained using 4 sets of features. The model trained on TV based ACFs perform significantly better yielding a relative UAR improvement of 13.63\% compared to the next best model that was trained on MFCC based ACFs. Performance of Formant based ACFs and openSMILE features seem very close in terms of UAR. The chance level F1 scores for normal (N), moderate (M) and severe (S) classes were 0.16, 0.38 and 0.39 respectively.

\begin{figure}[t]
  \centering
  \begin{subfigure}{0.4\columnwidth}
    \includegraphics[width=\linewidth]{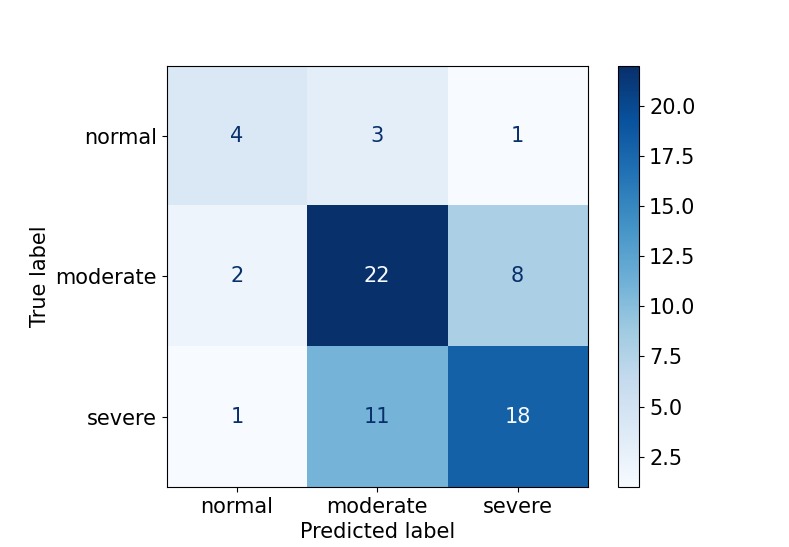} 
    \subcaption{Confusion Matrix}
    \label{fig:confmat}
    \end{subfigure}
    \begin{subfigure}{0.55\columnwidth}
    \includegraphics[width=\linewidth]{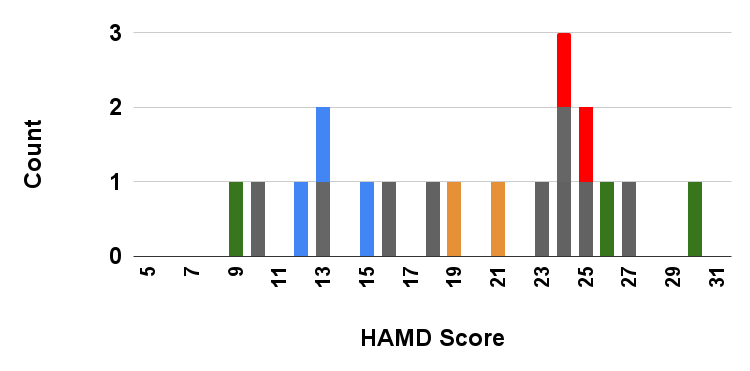} 
    \subcaption{HAMD of Misclassified Sessions}
    \label{fig:mis-hamd}
    \end{subfigure}
    \vspace{3pt}
    \caption{Analysis of Mis-classifications by TV based ACF Dilated CNN-LSTM Model}
  \label{fig:misclassifications}
\end{figure}

\vspace*{-5pt}
\begin{table}[h]
    \centering
    \small
    \caption{\textit{Overall Results of Segment-Level Classification}}
    \label{tab:overall-segment}
    \begin{adjustbox}{max width = \columnwidth}

\begin{tabular}{llrrl}
\toprule
\multicolumn{1}{c}{\textbf{Model}} & \multicolumn{1}{c}{\textbf{Features}} & \multicolumn{1}{c}{\textbf{Accuracy}} & \multicolumn{1}{c}{\textbf{UAR}} & \textbf{F1(N)/F1(M)/F1(S)} \\ \midrule
Dilated CNN & TV ACF & \textbf{50.84\%} & \textbf{0.5010} & \textbf{0.33 / 0.50 / 0.58} \\
Dilated CNN & MFCC ACF & 42.76\% & 0.4409 & 0.27 / 0.53 / 0.37 \\
Dilated CNN & Formant ACF & 38.22\% & 0.4089 & 0.26 / 0.42 / 0.40 \\
CNN & Opensmile & 40.20\% & 0.3980 & 0.21 / 0.42 / 0.48 \\ \bottomrule
\end{tabular}
    \end{adjustbox}
    \vspace{5pt}
\end{table}

\subsection{Session-Level Classification Results}
\label{sec:sess-resuls}

Using a low dimensional intermediate layer output of the segment-level classifier as en embedding, session-level RNNs were trained for all four feature sets. The overall session-level classification results can be found in Table \ref{tab:overall-session}. To benchmark the performance of the LSTM based session-level classifier, we used a conventional plurality voting approach where we used the mode of top 50\% of the segment level predictions based on the confidence as the session-wise prediction (ties were broken by randomly selecting a class out of the tied classes). This was done based on the hypothesis that highly confident segment-level predictions can produce a more reliable final prediction.

The LSTM based RNN model trained using TV based ACFs achieved a relative improvement of 27.47\% in UAR compared to the one trained using MFCC based ACFs. In general, the ACFs have yielded better UAR relative to the openSMILE features in session-level classification. Chance-level F1 scores for normal, moderate and severe classes for session-level classifications were 0.17, 0.39 and 0.38 respectively.
\vspace*{-5pt}
\begin{table}[h]
    \centering
    \small
    \caption{\textit{Overall Results of Session-Level Classification}}
    \label{tab:overall-session}
    \begin{adjustbox}{max width = \columnwidth}

    \begin{tabular}{llrrl}
    \toprule
\multicolumn{1}{c}{\textbf{Model}} & \multicolumn{1}{c}{\textbf{Features}} & \multicolumn{1}{c}{\textbf{Accuracy}} & \multicolumn{1}{c}{\textbf{UAR}} & \textbf{F1(N)/F1(M)/F1(S)} \\ \midrule
\multirow{4}{*}{\parbox{1.5cm}{LSTM based RNN}} & TV ACF & \textbf{62.86\%} & \textbf{0.5958} & \textbf{0.53 / 0.65 / 0.63} \\
 & MFCC ACF & 45.71\% & 0.4674 & 0.44 / 0.48 / 0.43 \\
 & Formant ACF & 37.14\% & 0.4028 & 0.32 / 0.4 / 0.36 \\
 & Opensmile & 37.14\% & 0.3729 & 0.21 / 0.39 / 0.43 \\ \hline
\multirow{4}{*}{\parbox{1.5cm}{Plurality Voting Classifier}} & TV ACF & 52.86\% & 0.5861 & 0.4 / 0.43 / 0.29 \\
 & MFCC ACF & 45.71\% & 0.5222 & 0.48 / 0.61 / 0.37 \\
 & Formant ACF & 30.00\% & 0.3784 & 0.63 / 0.12 / 0.22 \\
 & Opensmile & 42.86\% & 0.3243 & 0.0 / 0.42 / 0.51 \\ \bottomrule
\end{tabular}
    \end{adjustbox}
    \vspace{-10pt}
\end{table}

\section{Discussion}

\begin{figure}[t]
  \centering
    \includegraphics[width=0.55\linewidth]{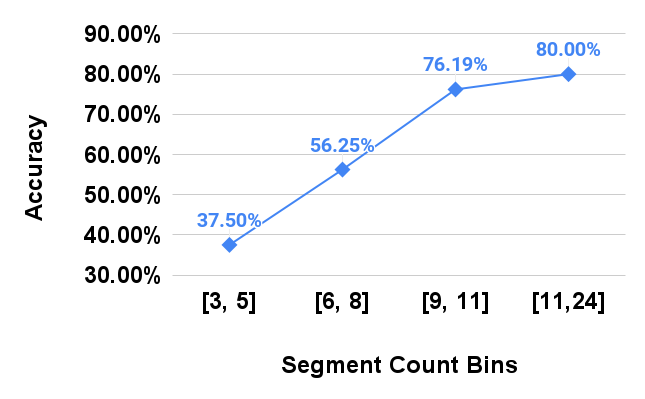} 
    \caption{Accuracy of session-wise classifications against the number of segments available for each session (TV based ACFs)}
  \label{fig:seg-count}
  \vspace{5pt}
\end{figure}

Based on the results reported in section \ref{sec:sess-resuls}, the ACFs derived from the TVs perform significantly better in this depression severity level classification task. This shows that using TVs as a direct measure for articulatory coordination can provide additional information in distinguishing different levels of depression severity. Our comparison among different feature sets helps to benchmark the performance of TV based ACFs with widely used other ACFs and eGeMAPS extracted using openSMILE.

The confusion matrix (Fig. \ref{fig:confmat}) obtained for the best performing model for the session-level classification shows that most of the confusions have occurred between adjacent classes (either normal-moderate or moderate-severe). We further analysed the HAMD scores of those sessions that were mis-classified between the moderate and severe classes (Fig. \ref{fig:mis-hamd}). There are instances where all or most of the sessions belonging to the same speaker (brightly-colored bars in the figure) are  misclassified. Out of the 19 misclassified sessions, 10 (3-green, 3-blue, 2-red, 2-orange) belonged to 4 speakers. The remaining 9 gray colored bars are from 9 different speakers. This suggests that a speaker dependent system may perform better than a speaker independent system by learning unique characteristics inherent to a particular speaker. 

The LSTM based RNN model for session-wise classifications is proposed to overcome the drawbacks in standard rule based approaches. A major drawback of these approaches is that there is no principled approach to choose rule thresholds. For instance, if we are using a plurality voting scheme which only considers $K$ most confident predictions and $K$ is chosen to maximize the metrics ($K_{max}$), the resultant scheme is biased towards the test set. Once $K$ is chosen, there is no guarantee that it will generalize to unseen test data. An alternative would be to pick the segment level predictions where the confidence is greater than a certain threshold $p_{th}$, however it's possible that there are no segment level predictions with better confidence than the threshold. With the proposed RNN based approach, the performance of the model is comparable with the result achieved using $K_{max}$ and the generalizability of the model is maintained. 

Looking at the performance of the segment-level and session-level classifiers it is evident that the strengths of the segment-level classifier are amplified as a result of its repeated usage in the session-level classifier. This also suggests the notion that the session level classifier is stronger on sessions with more segments. The notion is empirically verified by calculating the accuracy for ranges of segment counts in a session (Figure \ref{fig:seg-count}). 

\section{Conclusion and Future Work}
In this paper we proposed a new multi-stage architecture trained on TV based ACFs for depression severity classification which clearly outperforms the baseline models. We also established that the robustness of ACFs based on TVs holds beyond mere detection of depression and even in severity level classification.

This work can be extended to develop a multi-modal system that can take advantage of textual information obtained through Automatic Speech Recognition tools. Linguistic features can reveal important information regarding the verbal content of a depressed patient relating to their mental health condition.

\section{Acknowledgements}
\vspace{-5pt}
We would like to thank Dr. James Mundt for the depression databases MD-1\&2 \cite{MUNDT2007, MUNDT2012}. We also thank Dr. Thomas Quatieri and Dr. James Williamson for granting access to the MD-2 database \cite{MUNDT2012} which was funded by Pfizer. This work was supported by the UMCP \& UMB Artificial Intelligence + Medicine for High Impact Challenge Award and a Faculty Student Research Award from UMCP.

\bibliographystyle{IEEEtran}
\bibliography{references}
\end{document}